\RequirePackage{snapshot}
\documentclass[12pt, finaldraft, onecolumn]{IEEEtran}
\usepackage{graphicx}
\graphicspath{{./}, {./figs/}, {./figs/BCube_4Cases/}, {./figs/BCube_Hybrid/}, {./figs/BCube_Standard/}}
\DeclareGraphicsExtensions{.pdf,.jpeg,.png}
\usepackage[cmex10]{amsmath}
\usepackage{mathtools}
\usepackage{amssymb}
\usepackage{algpseudocode}
\usepackage{array}
\usepackage{amsmath}
\usepackage{amssymb}
\usepackage{graphicx}
\usepackage{array}
\usepackage{multirow}
\usepackage{rotating}
\usepackage{float}
\usepackage{color}
\usepackage{soul}
\usepackage{colortbl}
\usepackage{afterpage} \usepackage{setspace}
\usepackage{picins} 
\usepackage{randtext}
\catcode`\_=11\relax
\newcommand\email[1]{\_email #1\q_nil}
\def\_email#1@#2\q_nil{  \href{mailto:#1@#2}{{\randomize{#1}\emailampersat \randomize{#2}}}}

\newcommand\emailampersat{{\small@}} \catcode`\_=8\relax

\usepackage{doi}

\usepackage{bm}
\usepackage{hyperref}
\usepackage{breakurl} \usepackage{xcolor}
\hypersetup{
    colorlinks,
    linkcolor={red!50!black},
    citecolor={blue!50!black},
    urlcolor={blue!80!black}
}

\usepackage{spreadtab}
\usepackage{numprint}
\makeatother
\usepackage{cooltooltips}

\usepackage{caption}
\DeclareCaptionFormat{myformat}{#1#2#3\hrulefill}
\IEEEoverridecommandlockouts
\def\baselinestretch{1.6} 
\begin{document}
\title{Dynamic Bandwidth-Efficient BCube Topologies for Virtualized Data Center Networks}
\author{Vahid Asghari, Reza~Farrahi~Moghaddam and Mohamed~Cheriet
\thanks{The authors thank the NSERC of Canada for their financial support under Grant CRDPJ 424371-11 and also under the Canada Research Chair in Sustainable Smart Eco-Cloud.}}

\author{\IEEEauthorblockN{\small 
Vahid ASGHARI, 
Reza FARRAHI MOGHADDAM, 
and Mohamed CHERIET}\\
\IEEEauthorblockA{Synchromedia Lab and CIRROD, ETS (University of Quebec)}\\
\IEEEauthorblockA{Montreal, QC, Canada H3C 1K3}\\
\IEEEauthorblockA{Email: \email{vahid@emt.inrs.ca}, \email{imriss@ieee.org},
 \email{mohamed.cheriet@etsmtl.ca} } 
}

\maketitle
\thispagestyle{empty}
\begin{abstract}
\noindent Network virtualization enables computing networks and data center (DC) providers to manage their networking resources in a flexible manner using software running on physical computers. In this paper, we address the existing issues with the classic DC network topologies in virtualized environment, and investigate a set of DC network topologies with the capability of providing dynamic structures according to the service-level required by the active traffic in a virtual DC network. In particular, we propose three main approaches to modify the structure of a classic BCube topology as a topology benchmark, and investigate their associated structural features and maximum achievable interconnected bandwidth for different routing scenarios. Finally, we run an extensive simulation program to check the performance of the proposed modified topologies in a simulation environment which considers failure analysis and also traffic congestion. Our simulation experiments, which are consistent to our design goals, show the efficiency of the proposed modified topologies comparing to the classic BCube in terms of bandwidth availability and failure resiliency.
\end{abstract}
\newpage
\def\baselinestretch{1.8}
\section{Introduction}
\label{sec:introduction}
An efficient resource management is of great importance in future networks. This is especially the case in data center (DC) networks which would utilize software-defined networking (SDN) to improve resource and traffic management \cite{SurveySDN2014}.
The concept of SDN makes a separation between the control decisions (the control plane) and forwarding devices (the data plane) in a network. Recently, SDN draw much attention from both academia and industry by enabling innovative and adaptive network management systems \cite{SezerSDN2013,GothSDN2011}.
In typical SDN implementations such as that of OpenFlow, the network decision making process is logically centralized in software-based controllers and the network devices are considered as simple packet forwarding devices that can be programmed using an interface \cite{SurveySDN2014,OpenFlow2008}.

In recent years, DCs have evolved from passive elements in computing networks to become an active parts that dynamically adapt to different application requirements and manage their energy efficiency in large scale DC networks \cite{Koomey2011,NotEasyBeingGreen2012,Farrahi2014b}.
In this regard, the concept of virtualized DC network is proposed as a solution to provide better energy efficiency, resource utilization, management, and scalability \cite{Kulseitova2013}.
For instance, in \cite{Beloglazov2010}, an energy efficient resource management system is proposed for virtual DC networks which satisfies the required quality of service (QoS) and reduces the operational costs.

However, it has been observed that there has been much focus on improving the efficiency in computing and cooling systems (representing almost 70\% of a DC's total energy consumption), while less attention has been paid to the infrastructure of the DC network, which could account for a considerable amount of up to 20\% of the total energy consumption, as reported in 2006 \cite{BariSurvay2013}.
In this context, it has been shown in \cite{AlFares2008} that by interconnecting commodity switches in a DC network architecture, it is possible to provide a scalable interconnection bandwidth while reducing the architecture cost.
In \cite{Elastictree2010}, considering a Fat tree topology in a DC network, a network-wide energy optimizer was investigated such that it can dynamically optimize the power consumption at the DC network while satisfying the current traffic conditions.
In fact, for a given network topology and input traffic in a DC network, the proposed optimizer was set to find a minimum-power network subset which satisfies current traffic conditions and simply turns off those network components that are not needed under that given traffic condition. As a result, the proposed approach yielded 38\% energy savings under varying traffic conditions.
In some similar work, such as \cite{Honeyguide2012}, server virtualization techniques were used to increase the number of network active components that can be turned off in some traffic conditions, and consequently, increase the overall energy efficiency of the DC network.

In this paper, we investigate a group of DC network topologies with the capability to provide dynamic structures according to the service-level required by the active traffic in a virtual DC network. In particular, we propose certain modifications to the classic network structures to make them more dynamically adaptive to the requested bandwidth by a given DC traffic.
In this regard, we first consider a classic BCube topology as a benchmark for our analysis, and investigate its features regarding the network structure and interconnection bandwidth (IBW). Then, we propose the three main approaches to modify a classic BCube, namely, Horizontally, Vertically and Hybrid, and investigate their associated structural topology features and maximum achievable IBW for the cases of single- and multi-path routing scenarios.
We further define performance metrics based on the physical network topology characteristics and also the achievable bandwidth of the network with multipath routing protocol.
Finally, we test the performance of the proposed modified topologies in a simulation environment by running an extensive simulation program which  considers the failure analysis and also traffic congestion.

Herein, it is worth to mention that we have chosen the BCube topology as a base topology in our study since is currently considered by many DC designers toward disaggregated approaches to DCs. However, the contribution of this paper is not limited to a special topology or traffic, and the proposed approach is quite generic and can be easily modified and applied to any other network topology structure.

\section{Modified BCube Topologies for Bandwidth Efficiency}
\label{sec:MDCs}
In this section, we present the BCube DC network topology which is one of the well-known topologies used in the practical DC networks such as Sun's Modular DC \cite{SUNMDC}, HP's POD \cite{HPPOD}, and IBM's Modular DC \cite{IBMMDC}.
The BCube topology construction have been firstly proposed in \cite{Guo2009} as a server-centric approach, rather than the traditional switch-oriented practice. The general purpose of proposing such topology design is to `scale-out,' which means using multiple commodity switches in place of a single high-end switch. The goal of this design pattern is to reduce the cost of the network, since commodity switches are inexpensive and this makes BCube an appropriate design for large scale DC networks \cite{FarrahiFailure2014}.

As can be understand from the definition of a classic BCube topology, there are some restrictions on the size of the DC networks in terms of the number of servers and also the number of switches' ports.
For instance, in a classic BCube topology, the number of servers in each cube are restricted to be equal to the number of ports of switches. Another restriction is that there is no direct link between two switches or servers in a BCube construction.
Although these constraints are acceptable for small scale DCs with a normal traffic, they can impose considerable restrictions to the resource availability and management of large DC networks when multi-level QoS requirements are targeted.
Accordingly, to resolve this problem in the structure of the classic BCube topology, we propose a modified form of BCube topology, by utilizing available high-speed 10~Gbps ports in commodity switches to ultimately provide multiple levels of interconnected bandwidth in a DC network.

In the following, first we explain the classic BCube topology as a benchmark. Later, we explain our proposed modifications to the BCube topology and evaluate them in terms of features such as increase in the bandwidth efficiency and also failure resiliency.
\begin{figure}[tbh!]
\begin{tabular}{cc}
\includegraphics[width=9cm]{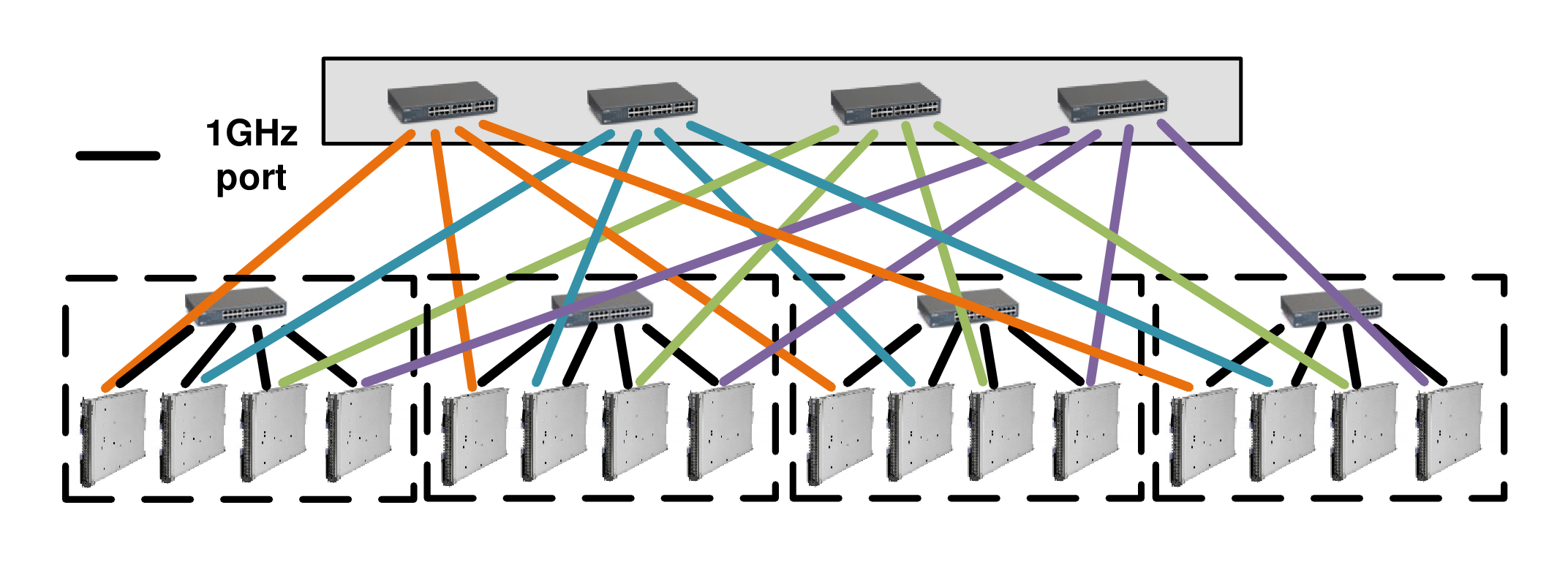} &
\includegraphics[width=9cm]{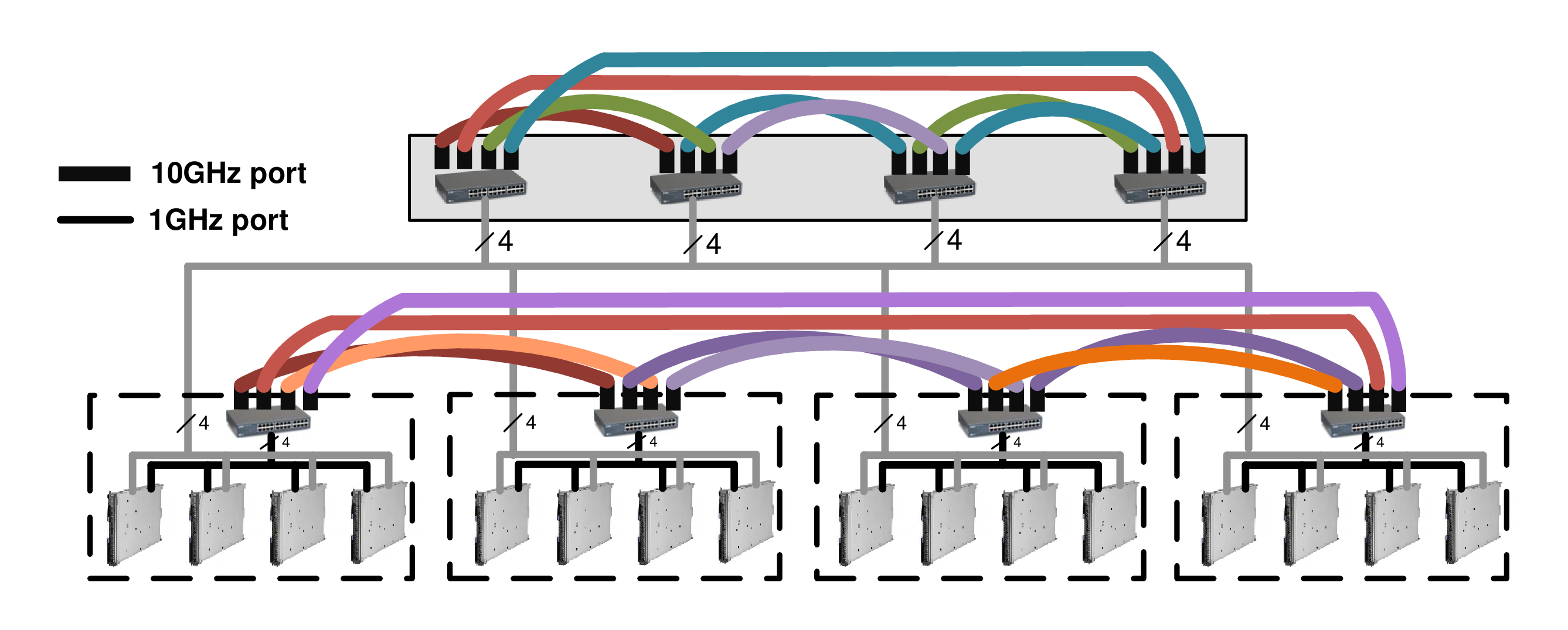} \\
(a) & (b) \\
\includegraphics[width=9cm]{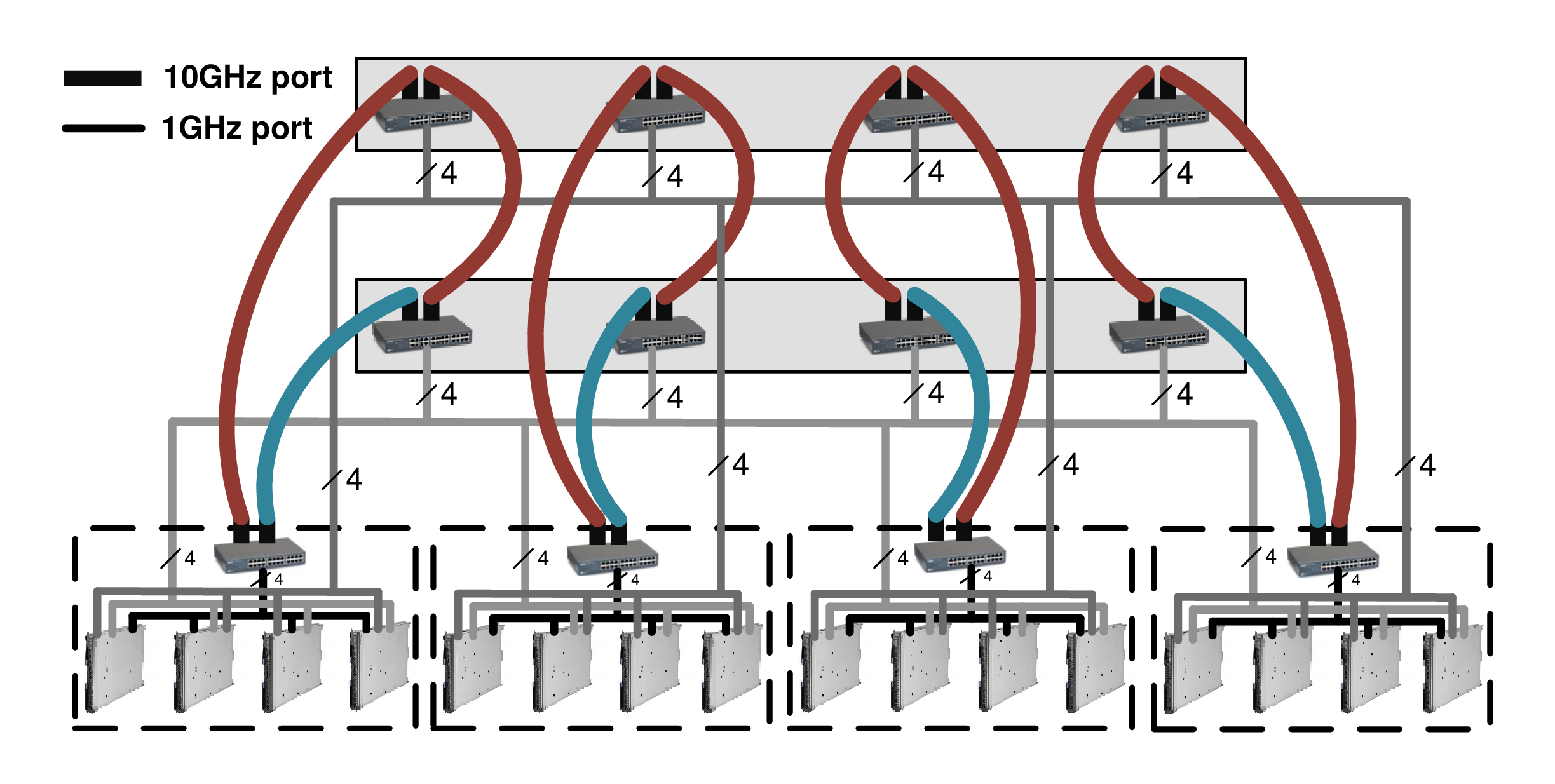} &
\includegraphics[width=9cm]{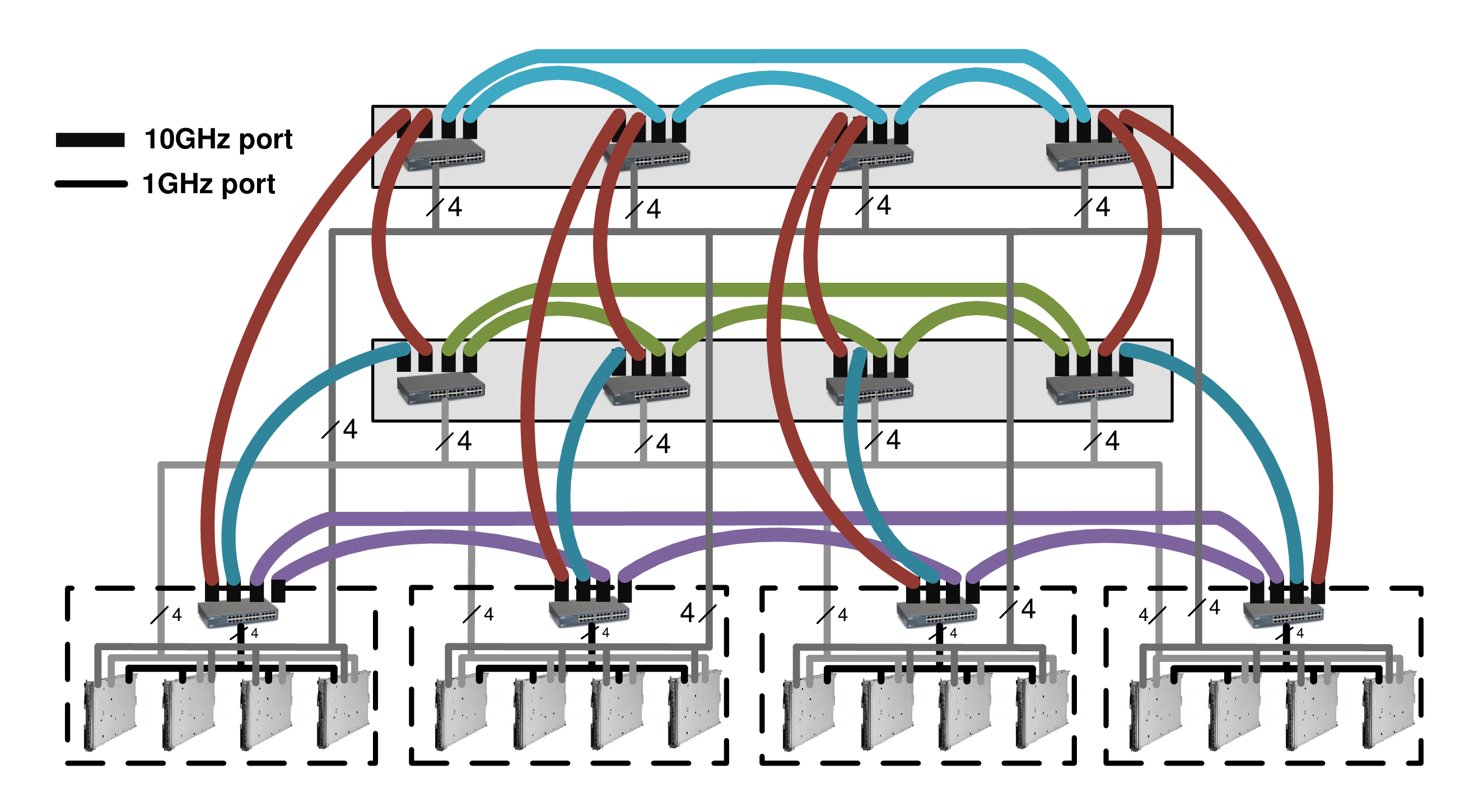} \\
(c) & (d) \\
\end{tabular}
\caption{Two illustrative samples of a) BCube with $k_{\rm{1G}}=4$ and $k_{\rm{s}}=2$, b) Horizontal-BCube with $k_{\rm{1G}}=4$, $k_{\rm{10G}}=4$ and $k_{\rm{s}}=2$, c) Vertical-BCube with $k_{\rm{1G}}=4$, $k_{\rm{10G}}=2$ and $k_{\rm{s}}=3$ and d) Hybrid-BCube with $k_{\rm{1G}}=4$, $k_{\rm{10G}}=4$, $k_{\rm{10G}}^{\rm{H}} = k_{\rm{10G}}^{\rm{V}} = 2$, and $k_{\rm{s}}=3$, topologies. Note that for the clarity of presentation in (b)-(d), the 1Gb/s connections are shown in a bundle form.}
\label{Fig-BcubeBCubeStar-schematic}
\end{figure}

\subsection{Classic BCube Topology}
\label{ssec:fattree_topo}
As shown by an example in Fig. \ref{Fig-BcubeBCubeStar-schematic}-(a), the classic BCube topology has a recursive structure, in which different levels of connectivity can be recognized between the switches and the servers. These levels in a BCube structure are shown as BCube~(level-$i$), where $i$ is the number of levels. A BCube~(level-0) is simply composed of $K$ servers connected to a single switch with $K(=\rm{PSwc})$ number of ports. A BCube~(level-1) is constructed from $N$ number of BCubes~(level-0) and $N$ number of switches with $K$ ports, where $K$ denote as the number of servers in each BCube~(level-0).
Each switch is connected to all $N$ BCubes~(level-0) sub-networks through its connection with one server of each BCube~(level-0).
Continuing the same steps, the general structure for the classic BCube~(level-$L$) topology is composed of $N$ number of BCubes~(level-$L-1$) and $N$ number of switches with $K$ ports. Again, each switch of BCube~(level-$L$) is connected with one server of each BCube~(level-0). It is worth to mention that each server in a classic BCube~(level-$L$) has $L + 1(=\rm{PSrv})$ ports, which are connected to level-$0$ to level-$L$ switches. Moreover, a classic BCube (level-$L$) consists of $N\times K$ number of servers and $N\times (L+1)$ number of switches, with each having $K$ number of ports. Fig. \ref{Fig-BcubeBCubeStar-schematic}-(a) shows the design for a classic BCube~(level-1) with $K=4$ and $N=4$.
\subsection{Proposed Modified BCube Topologies}
\label{subsec:star-topo}
In practice, one of the most important features in the design of DC networks is the network interconnection bandwidth.
For a given network topology, the interconnection bandwidth (IBW) is defined as the achievable speed at which the information can be exchanged between two servers. Moreover, the average network bandwidth value is nowadays considered as a parameter for the analysis of different topologies in DC networks. In particular, the network IBW can provide some useful insights along with failure and latency analysis in DC networks, while investigating the maximum number of link failures in which a network can tolerate before being split into multiple connected-component networks.
In this section, our aim is to present a few design approaches for maximizing the network IBW values in design of a BCube DC network topology.

The classic BCube construction makes sure that servers only connect to switches at different levels and they never connect directly to other servers. Similar rule considers between switches, i.e., switches never connect to other switches. By looking at the classic BCube topology in Fig. \ref{Fig-BcubeBCubeStar-schematic}-(a), we can treat the switches as some intermediate relays in which allow server-to-server communication. Therefore, it can be seen that the classic BCube topology suffers from low IBW values between different switch nodes in a network. It is important to note that considering the aforementioned feature in classic BCube topologies, this can affect the multiple-path flow between different nodes in a DC network, i.e., server-to-server (srv-srv), switch-to-switch (swc-swc), and server-to-switch (srv-swc) communication links.\footnote{Note that srv and swc stand for a server and a switch, respectively.}

On the other hand, the recent commercial switches provide lots of Gigabit ports and also several number of high speed 10~Gbps ports. It is worth mentioning that the classic BCube topology only uses the Gigabit ports. With our proposed modifications to the classic BCube construction, the high speed 10~Gbps ports are used to connect different level switches of classic BCube topology.
In particular, we propose using the high speed ports of switches in different layers (rows) of the modified BCube topology in various forms of \emph{Horizontally}, \emph{Vertically}, and \emph{Hybrid}, i.e., combination of horizontally and vertically, swc-swc connections.
Therefore, these high speed ports provide some additional links between switches in a BCube topology without requiring any extra switch or router.
The proposed modified BCube topologies simply addresses the low IBW values between switch nodes in classic BCube topology. It also connects directly the high speed ports of the switches at only the cost of cabling.

In the following, we formally explain the proposed modified BCube topologies, namely, Horizontal, Vertical and Hybrid, and then provide some comparisons among these topologies in terms of some common features. In our formulations, $k_{\rm{s}}(=\rm{PSrv})$, $k_{\rm{1G}}(=\rm{PSwc})$ and $k_{\rm{10G}}$ denote the number of server ports, the number of 1 Gigabit and high-speed 10 Gigabit switch ports, respectively.

For the case of Horizontal-BCube topology, the high speed ports of the switches on a given layer in a classic BCube are connected in a horizontal manner.
An example of a Horizontal-BCube Topology is shown in Fig. \ref{Fig-BcubeBCubeStar-schematic}-(b) for topology parameters of $k_{\rm{1G}} = 4$ and $k_{\rm{s}} = 2$. Note that in this example, the number of high speed 10~Gbps ports of the switches, $k_{\rm{10G}}$, is also set to $4$.
In contrast, for the case of Vertical-BCube topology, the high-speed switch ports of a given layer in a classic BCube are connected vertically to the high-speed ports of the overhead layer switches. An example of a Vertical-BCube topology is shown in Fig. \ref{Fig-BcubeBCubeStar-schematic}-(c) for topology parameters as $k_{\rm{1G}} = 4$, $k_{\rm{10G}} = 2$ and $k_{\rm{s}} = 3$.
Finally, for the case of Hybrid-BCube topology, the high-speed ports of the switches are divided into two sets of ports as $k_{\rm{10G}}^{\rm{H}}$ and $k_{\rm{10G}}^{\rm{V}}$, such that $k_{\rm{10G}} = k_{\rm{10G}}^{\rm{H}} + k_{\rm{10G}}^{\rm{V}}$, and each set of high speed ports are connected to horizontally and vertically, respectively, as explained for the Horizontal- and Vertical-BCube topologies. An example of a Hybrid-BCube topology is shown in Fig. \ref{Fig-BcubeBCubeStar-schematic}-(d) for topology parameters as $k_{\rm{1G}} = 4$, $k_{\rm{10G}} = 4$, $k_{\rm{10G}}^{\rm{H}} = k_{\rm{10G}}^{\rm{V}} = 2$, and $k_{\rm{s}} = 3$.
\subsubsection{Structural Properties}
\label{Structural_Prop}
Herein, our aim is to investigate the topological features of the proposed modified BCube topologies, namely, Horizontal, Vertical, and Hybrid BCubes. Study of these properties can help us to better understand the impact of the designing parameters on the performance of the proposed modified BCube topologies.

First, we calculate the number of additional links in Horizontal-BCube topology compare to the classic BCube topology.
In this regard, note that the number of cubes in a BCube topology is equal to the number of 1 Gigabit switch ports, $k_{\rm{1G}}$. Also the number of BCube levels is equal to the number of server ports, $k_{\rm{s}}$. Now, considering that there is always one link between two ports, we can obtain the number of additional horizontal links in a modified Horizontal-BCube compare to the classic BCube according to $k_{\rm{s}} ~ \left( \displaystyle{{k_{\rm{1G}}\ k_{\rm{10G}}}/{2}} \right)$. However, it is clear that the number of additional vertical links in Horizontal-BCube comparing to the classic BCube topology is zero. In this context, following the same approach used for the Horizontal-BCube topology, we can obtain the number of additional horizontal and vertical links for the Vertical- and Hybrid-BCube topologies as shown in Table \ref{table:T1}.
\begin{table}[ht]
\caption{Additional number of links in modified BCube Topologies compared to classic BCube topology.}    \centering   \begin{tabular}{c | c |c |c }   \hline  \hline   & Horizontal BCube & Vertical BCube & Hybrid BCube\\
 &  & & \scriptsize{$k_{\rm{10G}} = k_{\rm{10G}}^{\rm{H}} + k_{\rm{10G}}^{\rm{V}}$} \\ [1ex] \hline   Additional number of horizontal links  & $k_{\rm{s}}  \left( {k_{\rm{1G}} \ k_{\rm{10G}}}/{2} \right)$ & $0$ &  $k_{\rm{s}}  \left( {k_{\rm{1G}} \ k_{\rm{10G}}^{\rm{H}} }/{2} \right)$    \\ [1ex]
\hline Additional number of vertical links  & $0$ & $k_{\rm{1G}} \left( {k_{\rm{1G}} \ k_{\rm{10G}}}/{2} \right)$ &  $k_{\rm{1G}} \left( {k_{\rm{s}} \ k_{\rm{10G}}^{\rm{V}} }/{2} \right)$    \\ [1ex]
\hline \end{tabular}
\label{table:T1}
\end{table}

Considering the BCube topology properties presented in \cite{Guo2009}, it can easily be shown that the diameter of a classic BCube topology which is defined as the longest shortest path between all the server pairs is $4$. It can further be shown that for a classic BCube topology, the longest shortest path between two switches as well is $4$, and that of between a pair of one server and one switch is $3$.
However, using the same definition for the case of proposed modified BCube topologies, we obtain the same numbers for the longest shortest path between any server pairs, switch pairs and a pair of one server and one switch, as shown in Table \ref{table:T2}. It is worth to mention that for the most common case of $k_{\rm{s}}<4$ in Vertical and Hybrid topologies, the longest shortest path between switch pairs and a pair of one server and one switch are always equal to $3$.
\begin{table}[ht]
\caption{Length of longest shortest path in BCube Topologies.}    \centering \begin{tabular}{c | c |c |c |c}   \hline  \hline Between & Classic BCube & Horizontal BCube & Vertical BCube & Hybrid BCube \\[1ex] \hline  srv $\leftrightarrow$ srv  & $4$ & $4$ &  $4$  & $4$  \\[1ex]
swc $\leftrightarrow$ swc  & $4$ & $4$ &  $4~(3)^*$  & $4~(3)^*$  \\[1ex]
srv $\leftrightarrow$ swc  & $3$ & $3$ &  $3$  & $3$  \\ [1ex]
\hline  \end{tabular} \\[1ex]
Note: $^*$\small{For the most common case of $k_{\rm{s}}<4$.}
\label{table:T2}
\end{table}
\subsubsection{Bandwidth Properties}
\label{BW_Prop}
We further investigate the bandwidth efficiency in the proposed modified BCube topologies.
In this context, besides the utilization of single-path routing protocols in different networks, recent research has shown interests on multipath routing protocols for their effective role in fault tolerance and load balancing \cite{Meghanathan2010}.
Herein, we investigate the minimum-maximal IBW in the proposed topologies for both cases of single- and multi-path protocols.

First, we calculate the minimum-maximal single-path IBW between pair of network nodes in different BCube topologies.
In this case, since we only consider a single-path between any pair of nodes, it can be shown for the classic BCube topology structure that the minimum-maximal IBW between a pair of servers, a pair of switches and a pair of one server and one switch is equal to the bandwidth of a single 1G link, i.e., $B_{\rm{1G}}$. However, for the modified BCube topologies, we obtain that for the same level (horizontally and/or vertically) switch to switch connections, the minimum-maximal IBW value can be increased to the bandwidth of a single 10~Gbps link, i.e., $B_{\rm{10G}}$, as illustrated in Table \ref{table:T3}.
This can be interpreted such that using a single-path routing protocol, the proposed topology modifications to a classic BCube topology cannot generally affect the achievable IBW value between different pairs of network nodes, compared to the classic BCube topology.
It can, however, introduce a significant increase to the same level (horizontally and/or vertically) switches connections in the modified BCube topologies.
\begin{table}[ht]
\caption{Minimum maximal single-path bandwidth in BCube Topologies.}    \centering \begin{tabular}{c | c |c |c |c}   \hline  \hline Between & Classic BCube & Horizontal BCube & Vertical BCube & Hybrid BCube \\[1ex] \hline  srv $\leftrightarrow$ srv   & $B_{\rm{1G}}$ & $B_{\rm{1G}}$ &  $B_{\rm{1G}}$  & $B_{\rm{1G}}$  \\[1ex]
swc $\leftrightarrow$ swc  & $B_{\rm{1G}}$ & $B_{\rm{1G}}$ ($B_{\rm{10G}}$)$^*$ &  $B_{\rm{1G}}$ ($B_{\rm{10G}}$)$^{**}$ & $B_{\rm{1G}}$  ($B_{\rm{10G}}$)$^{*,**}$ \\[1ex]
srv $\leftrightarrow$ swc  & $B_{\rm{1G}}$ & $B_{\rm{1G}}$ &  $B_{\rm{1G}}$  & $B_{\rm{1G}}$  \\ [1ex]
\hline  \end{tabular} \\[1ex]
Notes: $^*$\small{For the same horizontal level switches.} $^{**}$\small{For the same vertical level switches.}
\label{table:T3}
\end{table}

Finally, for the case of using multi-path routing protocol, we calculate the minimum-maximal IBW between pair of network nodes in different BCube topologies. In fact, it is considered that any pair of nodes can be connected through multiple independent paths.
In this case, since each server has only $k_{\rm{s}}$ number of 1G ports, the achievable minimum-maximal IBW values between a pair of servers and between a pair of one server and one switch are equal to $k_{\rm{s}} \ B_{\rm{1G}}$, for all BCube topologies, as shown in Table \ref{table:T4}.
In order to calculate the achievable IBW between a pair of switches, we should note that each switch in a classic BCube topology has only $k_{\rm{1G}}$ number of 1G links. Accordingly, it can be obtained that the minimum-maximal IBW between a pair of switches in a classic BCube is equal to $k_{\rm{1G}} \ B_{\rm{1G}}$. However, for the case of modified BCube topologies, it is considered that each switch has also $k_{\rm{10G}}$ number of 10G links.
Considering the possible switch to switch connections in a Horizontal BCube topology, we can obtain the minimum-maximal IBW between a pair of switches by adding the amount of achievable bandwidth through all available 1G and 10G links, i.e., $k_{\rm{10G}} \ B_{\rm{10G}}+k_{\rm{1G}} \ B_{\rm{1G}}$.
From the structure of the proposed Vertical Bcube topology, it can be shown that, in addition to the $k_{\rm{10G}}$ number of 10G links, there are $(k_{\rm{1G}}-1)$ number of 1G links at each $(k_{\rm{s}}-1)$ number of switch levels (not including the chosen switch level). Accordingly, we can obtain the achievable minimum-maximal IBW value between a pair of switches according to $k_{\rm{10G}} \ B_{\rm{10G}} + (k_{\rm{s}}-1)(k_{\rm{1G}}-1) \ B_{\rm{1G}}$.
Considering the same approach, we can obtain the achievable minimum-maximal IBW between a pair of switches in a Hybrid Bcube topology as shown in Table \ref{table:T4}.
Our calculation for the case of multi-path routing protocol has shown a significant increase in the achievable IBW values between the switch nodes in the proposed modified BCube topologies compared to the classic BCube topology. This confirms the effectiveness of the proposed topology modification to a classic BCube topology for the case of using multi-path routing protocol.
However, it has been shown that the proposed topology modifications cannot affect the achievable IBW value between a pair of servers or between a pair of one server and one switch in a network.
\begin{table}[ht]
\caption{Minimum maximal multiple-path bandwidth in BCube Topologies.}    \centering \begin{tabular}{c | c |c |c |c}   \hline  \hline Between & Classic BCube & Horizontal BCube & Vertical BCube & Hybrid BCube \\[1ex] \hline  srv $\leftrightarrow$ srv   & $k_{\rm{s}} \ B_{\rm{1G}}$ & $k_{\rm{s}} \ B_{\rm{1G}}$ &  $k_{\rm{s}} \ B_{\rm{1G}}$  & $k_{\rm{s}} \ B_{\rm{1G}}$  \\[1ex]
swc $\leftrightarrow$ swc  & $k_{\rm{1G}} \ B_{\rm{1G}}$ & $k_{\rm{10G}} \ B_{\rm{10G}}$ &  $k_{\rm{10G}} \ B_{\rm{10G}}$  & $k_{\rm{10G}} \ B_{\rm{10G}}+k_{\rm{1G}} \ B_{\rm{1G}} $  \\
          &         &  $+ k_{\rm{1G}} \ B_{\rm{1G}}$   & $+ (k_{\rm{s}}-1)(k_{\rm{1G}}-1) \ B_{\rm{1G}}$  &    $+(k_{\rm{s}}-1)(k_{\rm{1G}}-1) \ B_{\rm{1G}}$           \\[1ex]
srv $\leftrightarrow$ swc  & $k_{\rm{s}} \ B_{\rm{1G}}$ & $k_{\rm{s}} \ B_{\rm{1G}}$ &  $k_{\rm{s}} \ B_{\rm{1G}}$  & $k_{\rm{s}} \ B_{\rm{1G}}$  \\ [1ex]
\hline  \end{tabular}
\label{table:T4}
\end{table}
\section{Performance Metrics}
\label{sec:Tanh}
In our simulation analysis, we first investigate the topology characteristics in the proposed modified BCube topologies when they are used as the  physical layer of the network. In particular, while considering the impact of component level failures in different BCube topology components, i.e., servers, switches and physical links, we investigate the metrics related to the network topology size such as maximum and average relative sizes of the network connected components.

In this case, using the same approach similar to those used for the analysis in \cite{Albert2000a,FarrahiFailure2014}, the relative size (RS) of the largest connected component in relation to the number of existent servers, ${\rm{RS}}_{\max}$, is given by
\begin{align}
{\rm{RS}}_{\max}\left( t_1 \right) & = & \mathop {\max}\limits_{i \in I} \left\{ {\rm{RS}}\left( t_1, i \right) \right\} = \frac{\mathop {\max}\limits_{i \in I} \left\{ n_{{\rm{srv}},i}\left( t_1 \right) \right\} }{\sum_{i\in I} {n_{{\rm{srv}},i}\left( t_1 \right)}},
\label{Metrics_1}
\end{align}
where $n_{{\rm{srv}},i}(t_1)$ denotes the number of servers available in the $i$-th connected components at time $t_1$.
In fact, ${\rm{RS}}_{\max}$ is defined as the maximum of the individual RSs of all connected components at time $t_1$.
Now, using the RS definition given in (\ref{Metrics_1}), we further define the maximum absolute RS ($\rm{ARS}_{\max}$) which is defined as the maximum of the individual ARSs of all connected components at time $t_1$ divided by the total number of available servers in all connected components at time $t_0$, according to
\begin{align}
{\rm{ARS}}_{\max}\left( t_1 \right) = \mathop {\max}\limits_{i \in I} \left\{ {\rm{ARS}}\left( t_1, i \right) \right\} = \frac{\mathop {\max}\limits_{i \in I} \left\{ n_{{\rm{srv}},i}\left( t_1 \right) \right\} }{N_{\rm{srv}}},
\label{Metrics_2}\
\end{align}
where $N_{\rm{srv}}$ denotes the total number of initial available servers in all the connected components at the start time $t_0$, i.e., $N_{\rm{srv}}=\sum_{i\in I}{n_{{\rm{srv}},i}\left( t_0 \right)}$.
Note that while the parameter ${\rm{RS}}_{\max}$ represents a RS value for the current state of the DC network, the parameter ${\rm{ARS}}_{\max}$ is defined such that it provides a relative size of the network based on the initial state of the network.
Although each of these two measures has its own importance in the design of DC networks, we only present results regarding the measure ${\rm{ARS}}_{\max}$ in this contribution.

We now calculate the achievable bandwidth of the network with the proposed BCube topologies for the case multi-path routing protocol.\footnote{For the case of single-path protocol, it is essential to implement an optimal routing algorithm. Note that this is beyond the scope of our contribution in this paper and we take this part into consideration in our future works in his direction.} In our simulation, we also consider a component level failure to investigate the impact of failure on the achievable bandwidth between a pair of servers in different BCube topologies \cite{FarrahiFailure2014}.

In this case, we define the average minimum-maximal IBW of the network in the $i$-th connected components at time $t_1$ according to
\begin{align}
{\rm{IBW}} \left( t_1, i \right) = {\rm{average}} \left( \left\{ {\rm{IBW}} \left( {{\rm{srv}}_l,{\rm{srv}}_m} \right)  \Big{\vert} ~\forall~ l,m \in i  \right\} \right), \label{Metrics_3}\
\end{align}
where ${\rm{IBW}} \left( {{\rm{srv}}_l,{\rm{srv}}_m} \right)$ denotes the minimum-maximal IBW between two of the available servers in the $i$-th connected components. Note that using multi-path routing protocol, all the available paths are considered for communication between ${\rm{srv}}_l$ and ${\rm{srv}}_m$.

In the next section, a simulation analysis approach that generates, evaluates, and aggregates snapshots of a DC network along time considering the failure factor is used to evaluate the performance of the proposed BCube topologies in a data center network. The detailed of this approach can be found in \cite{FarrahiFailure2014}.
\section{Simulation Results}
\label{sec:Results}
In this section, some of previously-discussed results discussed are verified using a simulation platform. In order to have a realistic evaluation, the platform considers failure of network components along the time \cite{FarrahiFailure2014} in addition to network traffic congestion. In order to model traffic congestion, a congestion degree parameter $\gamma$ is introduced. When there is no traffic in the DC network, $\gamma=0$, a fully congested network has a $\gamma$ value of 1. For a specific $\gamma$ value, the congestion degree of every link is randomly determined from a truncated Gaussian distribution centered at $\gamma$. Having all links' bandwidth adjusted according to their individual congestion degree, the IBW is calculated following a snapshot-based Monte Carlo approach \cite{FarrahiFailure2014}.

\begin{figure}[!htb]
\centering
\begin{tabular}{cc}
\includegraphics[width=0.48\linewidth]{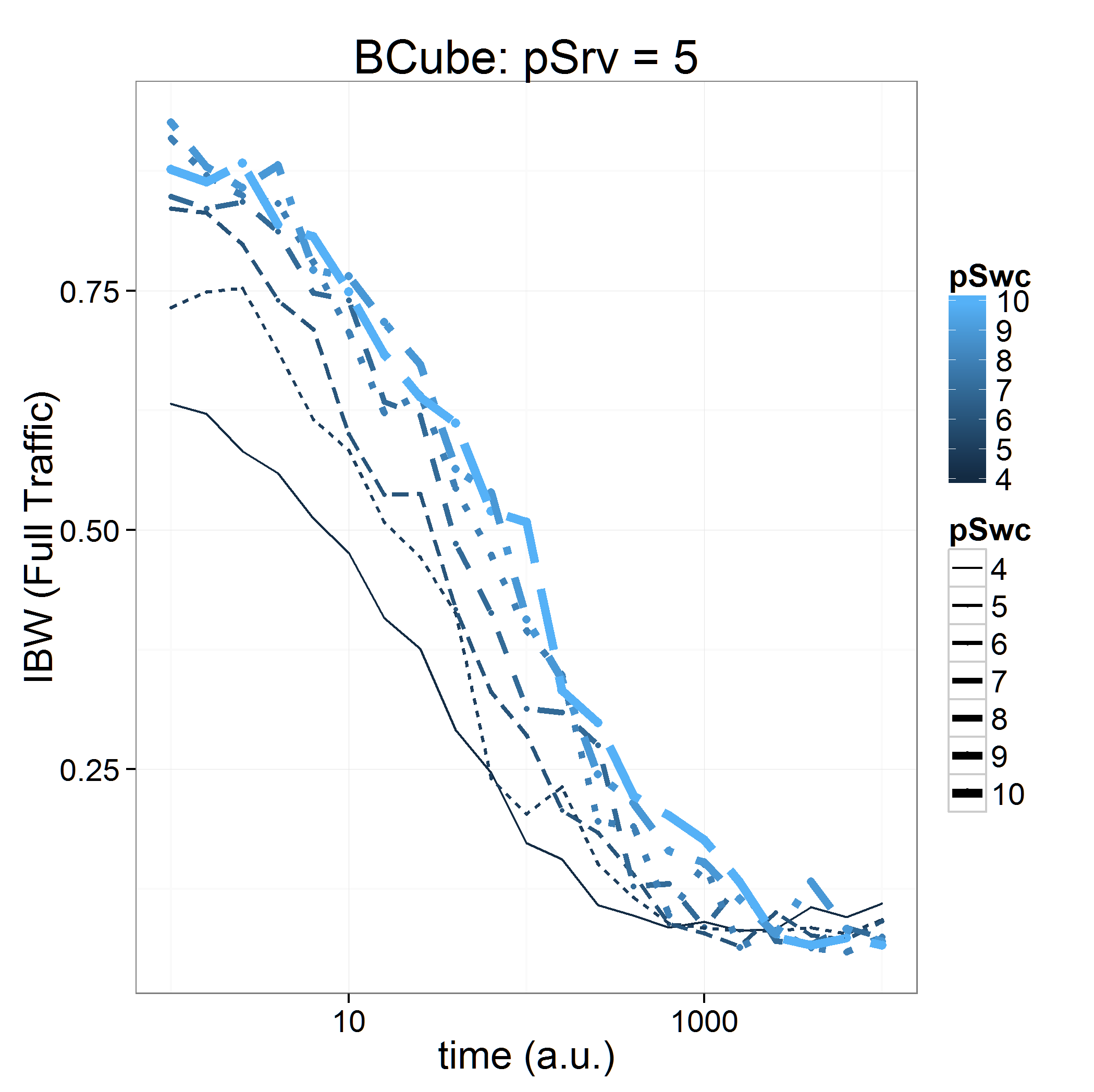} &
\includegraphics[width=0.48\linewidth]{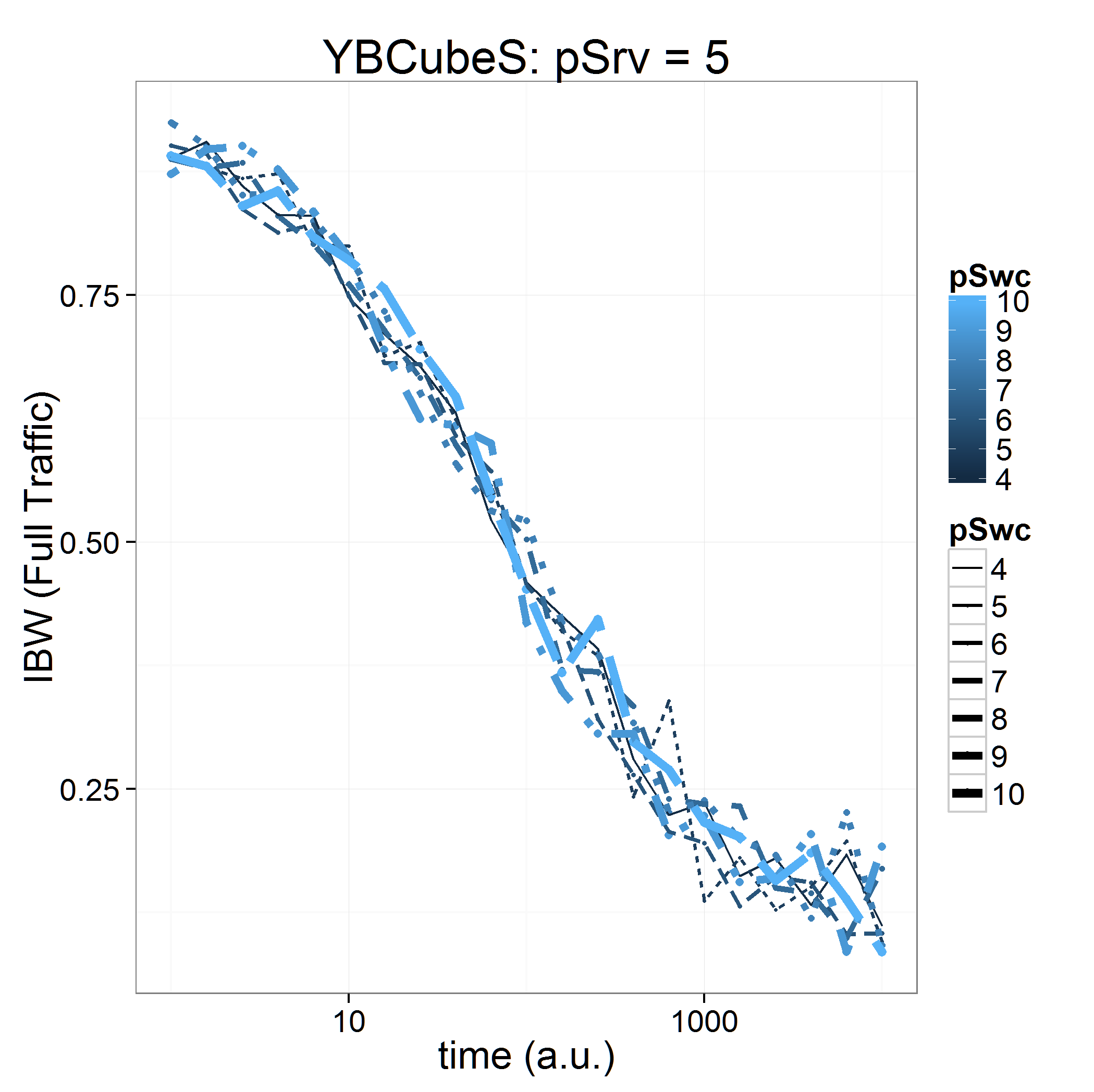} \\
(a) Classic BCube & (b)  Hybrid BCube
\end{tabular}
\caption{The IBW performance of the classic BCube topology versus the proposed Hybrid-BCube topology.}
\label{fig_bcube_class_hyb_traf1}
\end{figure}

First, the impact of traffic congestion on the performance of BCube topologies is illustrated in Fig. \ref{fig_bcube_class_hyb_traf1} in which a high-degree of congestion associated with $\gamma=0.95$ is studied. As can be seen from the figure, the classic BCube highly suffers from the congestion, especially when the number of 1G ports of switches is low. In contrast, the Hybrid-BCube topology shows a stable and robust behavior regardless of the $pSwc(=k_{1G})$ ports. It is worth mentioning that the decreasing behavior of the IBW along time is the direct result of the failure of components without any maintenance and replacement.

\begin{figure}[!htb]
\centering
\begin{tabular}{@{}c@{}@{}c@{}@{}c@{}}
\includegraphics[width=0.36\linewidth]{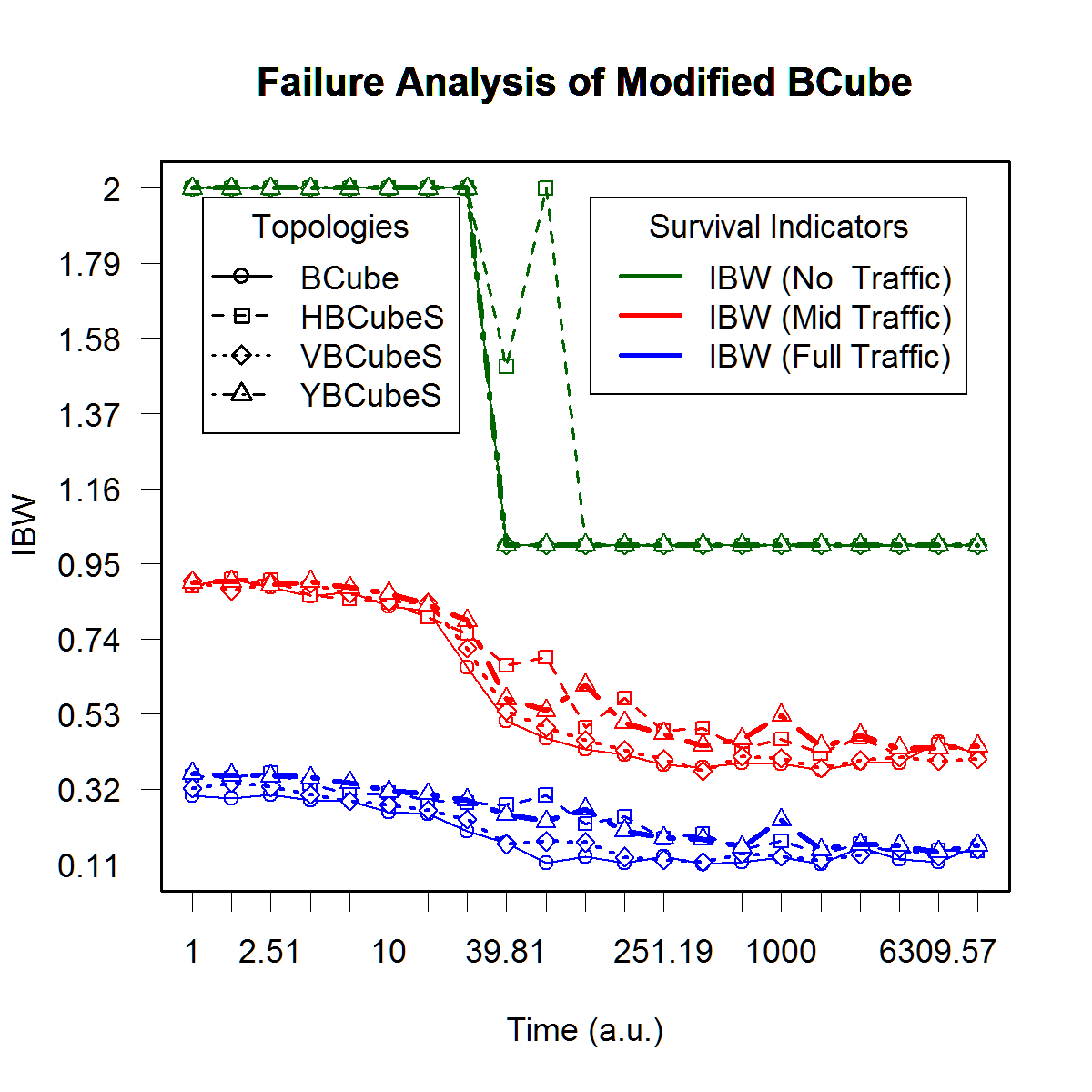} &
\includegraphics[width=0.36\linewidth]{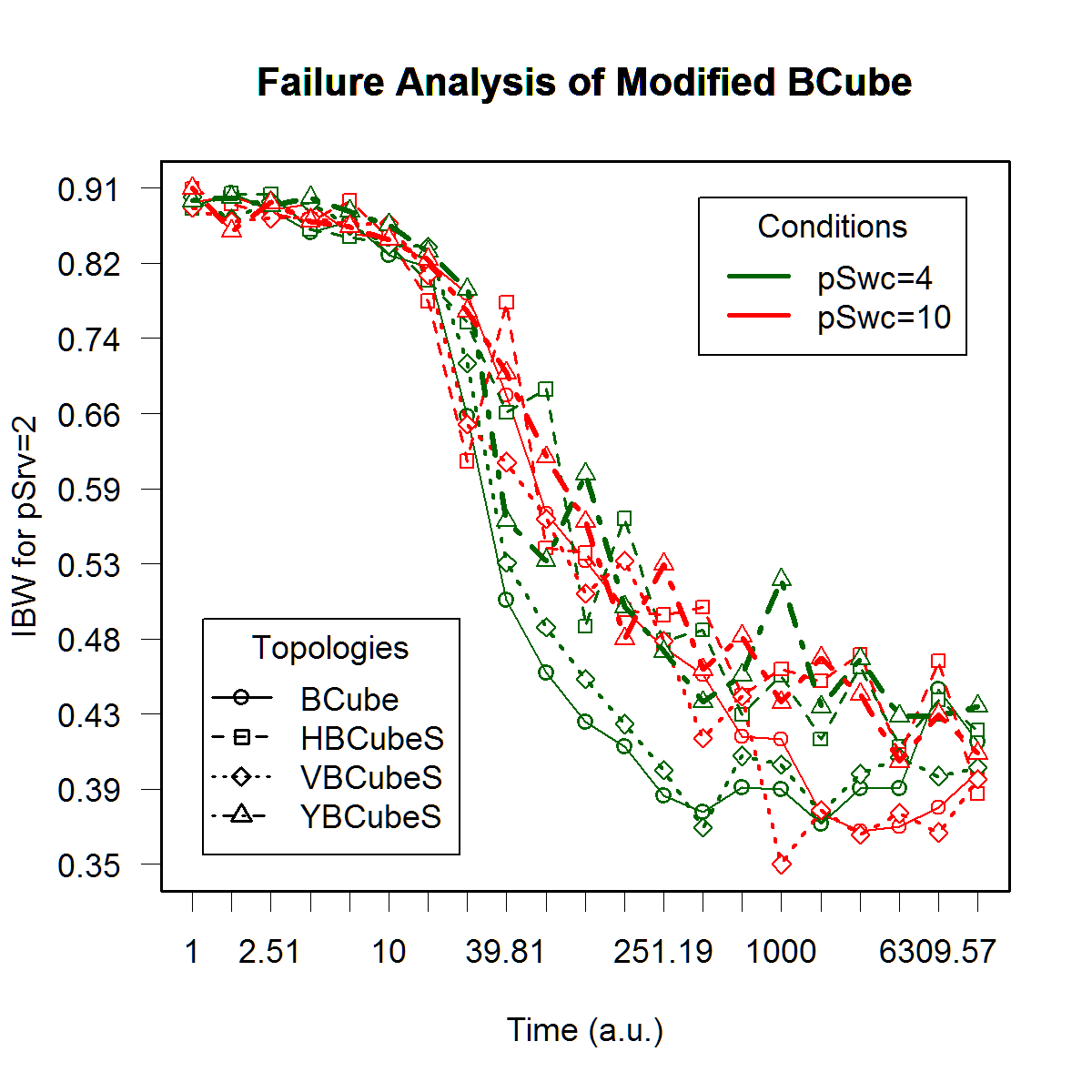} &
\includegraphics[width=0.36\linewidth]{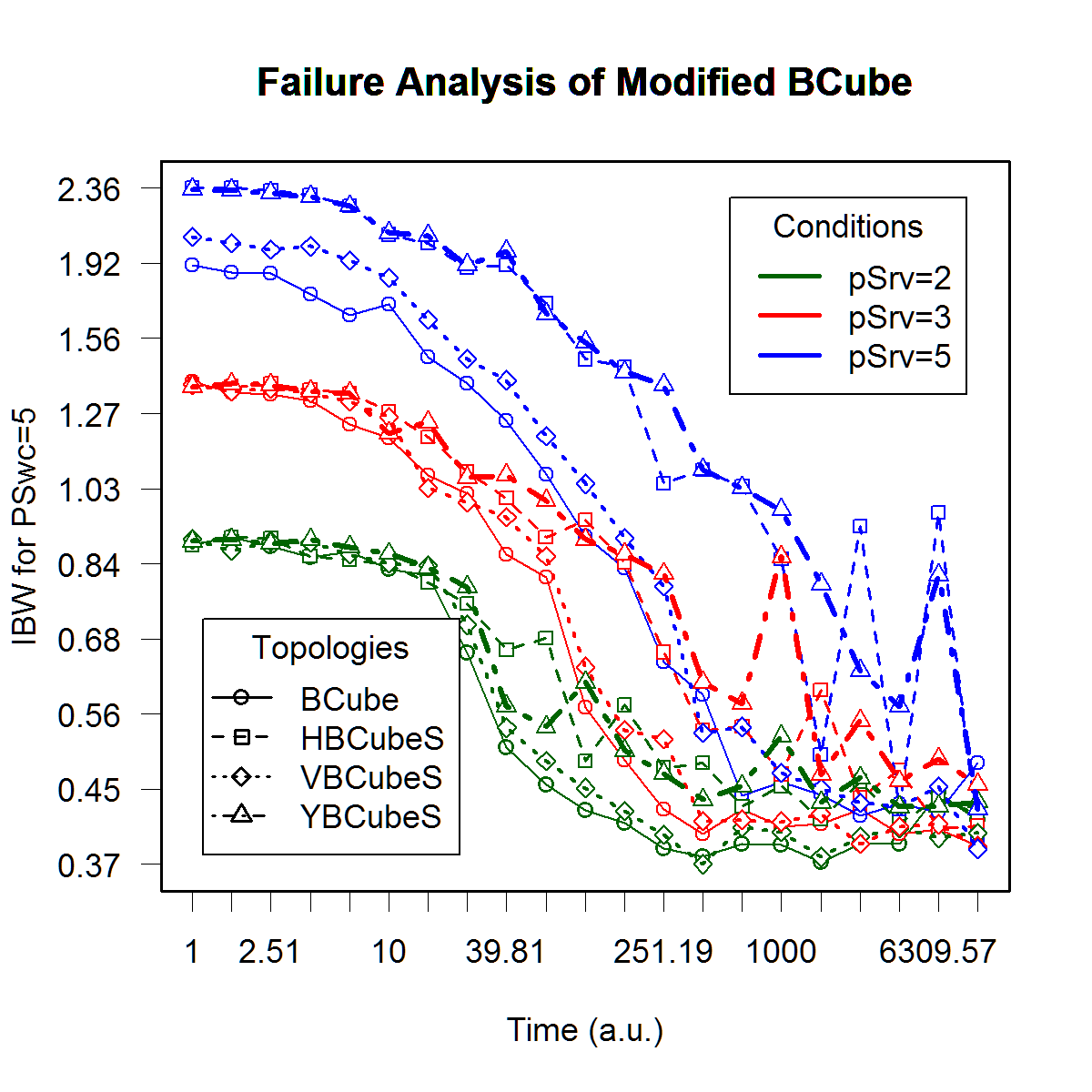} \\
(a) & (b) & (c)
\end{tabular}
\caption{The IBW performance of all the four BCube topologies.
a) The performance resiliency against traffic congestion and failure.
b) The impact of the $\rm{pSwc}$ on the performance.
c) The impact of the $\rm{pSrv}$ on the performance.}
\label{fig_bcube_4case1}
\end{figure}

Fig. \ref{fig_bcube_4case1} provides a comparison among all four BCube topologies. In Fig. \ref{fig_bcube_4case1}-(a), the impact of variation in $\gamma$ value on the IBW performance of the proposed topologies is shown. Three values for $\gamma$ is used: $\gamma=0.05, 0.5, 0.95$. As can be seen, both Horizontal- and Hybrid-BCube topologies show a better performance. As expected, with increase in the degree of congestion, the performance of all topologies is affected. In Fig. \ref{fig_bcube_4case1}-(b), the impact of the number of 1G ports of switches is illustrated. The case of low $\rm{pSwc}$ is a higher degree of degradation, especially for the classic and Vertical-BCube topologies. And, finally, in Fig. \ref{fig_bcube_4case1}-(c), the impact of the number of server's port, i.e., the number of BCube levels, is studied. As expected, with increasing $\rm{pSrv}$ and the number of levels, the performance of Horizontal- and Hybrid-BCube topologies is significantly improved. In particular, the time to degrade to half initial performance is extended by 10 folds when $\rm{pSrv}$ is increased from 2 to 5. Although classic and Vertical-BCube topologies also benefit from better IBW at time intervals near $t_0$, they show a very sensitive behavior with respect to failure as times go on, and all benefits of higher number of levels is diminished. In general, it is concluded that both Horizontal- and Hybrid-BCube topologies show a good performance, and even in some cases the Horizontal-BCube has a better performance. However, as will be discussed in the next section, the flexibility of Hybrid-BCube in providing more arbitrary topologies makes us to conclude that this topology is the first choice in designing physical layer of DC networks.

\section{Discussions}
\label{sec:Discussions}
So far, we illustrated the high resiliency of the proposed modified topologies with respect to both failure and also traffic congestion in terms of their high Interconnected BW and relative size values. However, we would like to take this opportunity and discuss some benefits of the proposed topologies in the design of the smart DC networks.

Generally, network virtualization techniques aim at decoupling the functions of a classic network providers into two key parts, i.e., the network infrastructure providers (InPs) and network service providers (SePs). InPs are considered to manage the physical infrastructure of the network, and SePs are considered to create virtual networks (VNets) by aggregating and sharing resources from multiple InPs and offer end-to-end network services \cite{ChowdhuryNetVir2009}. In fact, each VNet consists of an overlay of virtual nodes and links in which are a subset of the underlying physical network resources. It is worth noting that a VNet in a virtualization environment can specify a network topology based on its needs, such as having low latency for high-performance computing workloads or a high
interconnected BW for data processing workloads \cite{Webb2011}. In the following, our aim is to explain two new concepts of topology matching and topology on-demand in a virtual environment that could considerably reduce the overhead of virtual networks, and could also bring dynamic adaptability to variations in network requirements.
\subsection{Topology Matching: Beyond Topology Mapping in Virtual Networks}
Topology mapping is a well-known action in operating virtual network that allows {\em mapping} an arbitrary virtual network topology on the fix topology of the physical layer. In a virtualization environment, topology mapping stands for expressing a requested VNet topology sent to the service providers in terms of specific layout patterns of the interconnections of network elements, such as links and nodes, along with a set of specific service-oriented constraints, such as CPU capacity and link bandwidth \cite{Wang2011}. Various parameters are considered in designing a topology mapping algorithm. Despite the freedom that this approach provides, it could impose a considerable overhead in both mapping action itself and also in the operation of the network because of non-optimal solution obtained from a particular algorithm, especially when many VNets are mapped on the same physical layer. This disadvantage could be very significant in the case of big physical layer networks that are more popular in resource sharing approaches such as cloud computing.

In the topology mapping, there is always a downward mapping between the virtual layer elements and the physical layer elements. This one-way approach would result in cases in which the mapped topology on the physical layer and the original requested topology have a significant {\em topological} distance despite having a negligible distance in terms of requested service-oriented constraints. This disparity, which is the result of unawareness of the VNet about the actual physical topology, could result in a premature default of the service because of component failures or DC traffic congestion.

Herein, we propose considering the concept of {\em topology matching} instead of topology mapping. In the proposed topology matching approach, the topology of a VNet is the outcome of a negotiation and exchange interaction between the VNet, the SeP, and the InP. In such a process, a suggested topology from the Vnet would simply denied if the SeP/InP could not match it on the actual physical topology. The feedback that the InP provides to the Vnet helps them to converge to a working matchable topology that would perfectly fit on the actual DC topology.

Although the concept of topology matching seems very interesting, it could simply fail to provide any converged topology in many cases if the physical layer does not enough complexity and connectivity in terms of its topology. Motivated by the results obtained in this paper for the modified BCube topologies proposed in Sec. \ref{subsec:star-topo}, we believe this modifications to the physical network platforms, especially the Hybrid-BCube topology, are able to increase the chance of providing appropriate match topologies to a VNet request while conforming to the requested interconnection BW by the VNet.
\subsection{Topology on-Demand (ToD)}
A direct consequence of the vision of topology matching is in handling of dynamic changes in the VNet requirements. Many applications do not require a constant high bandwidth maintained between their nodes. However, they do need temporal, burst-like high-IBW in their operation along time. Such ephemeral requirements are usually addressed using dynamically changing the allocated bandwidth to the VNet network along the links of its topology.

\begin{figure}[tbh!]
\normalsize
\centering
\begin{tabular}{@{}c@{}|@{}c@{}}
\includegraphics[width=.52\linewidth]{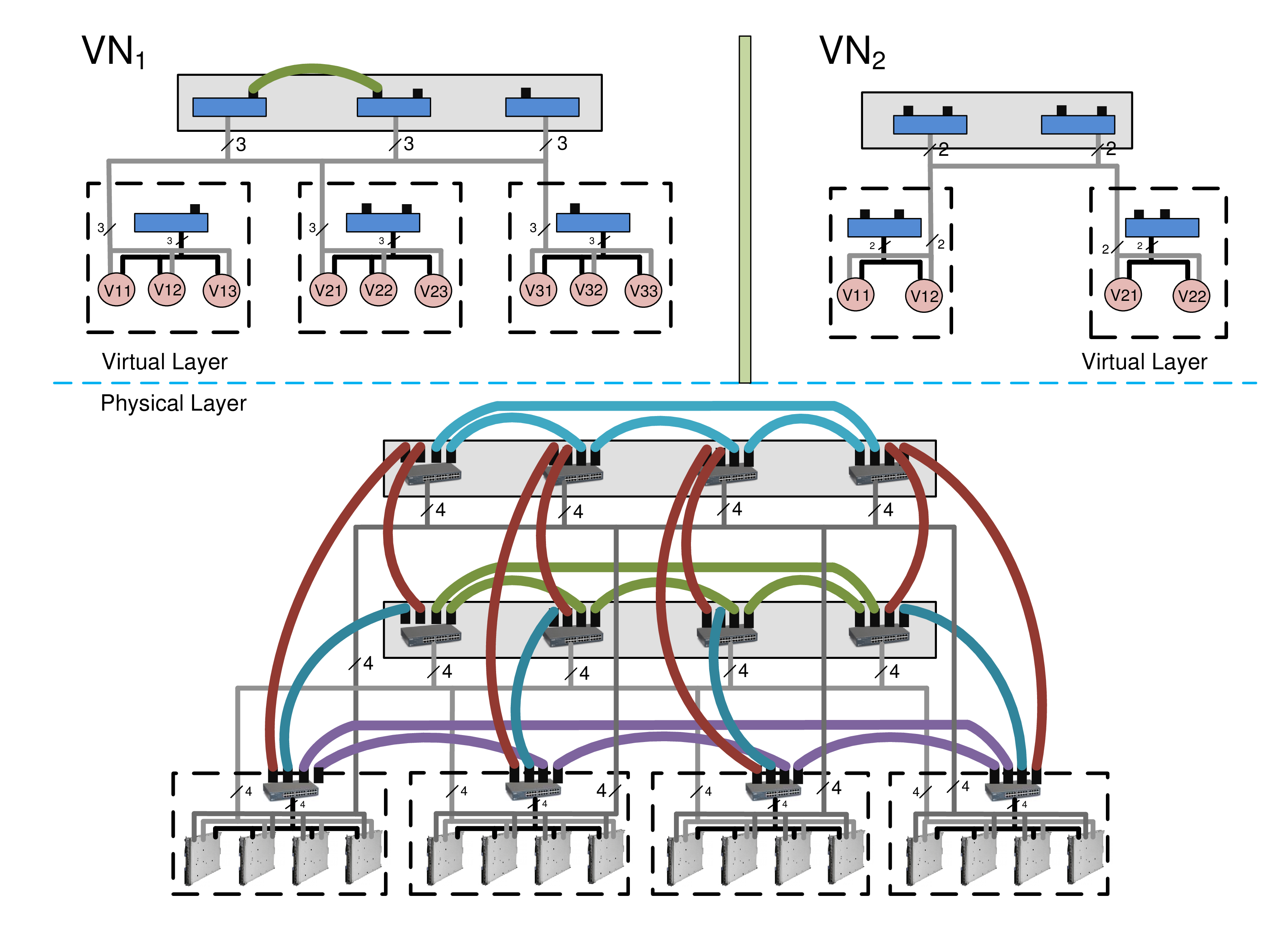} &
\includegraphics[width=.52\linewidth]{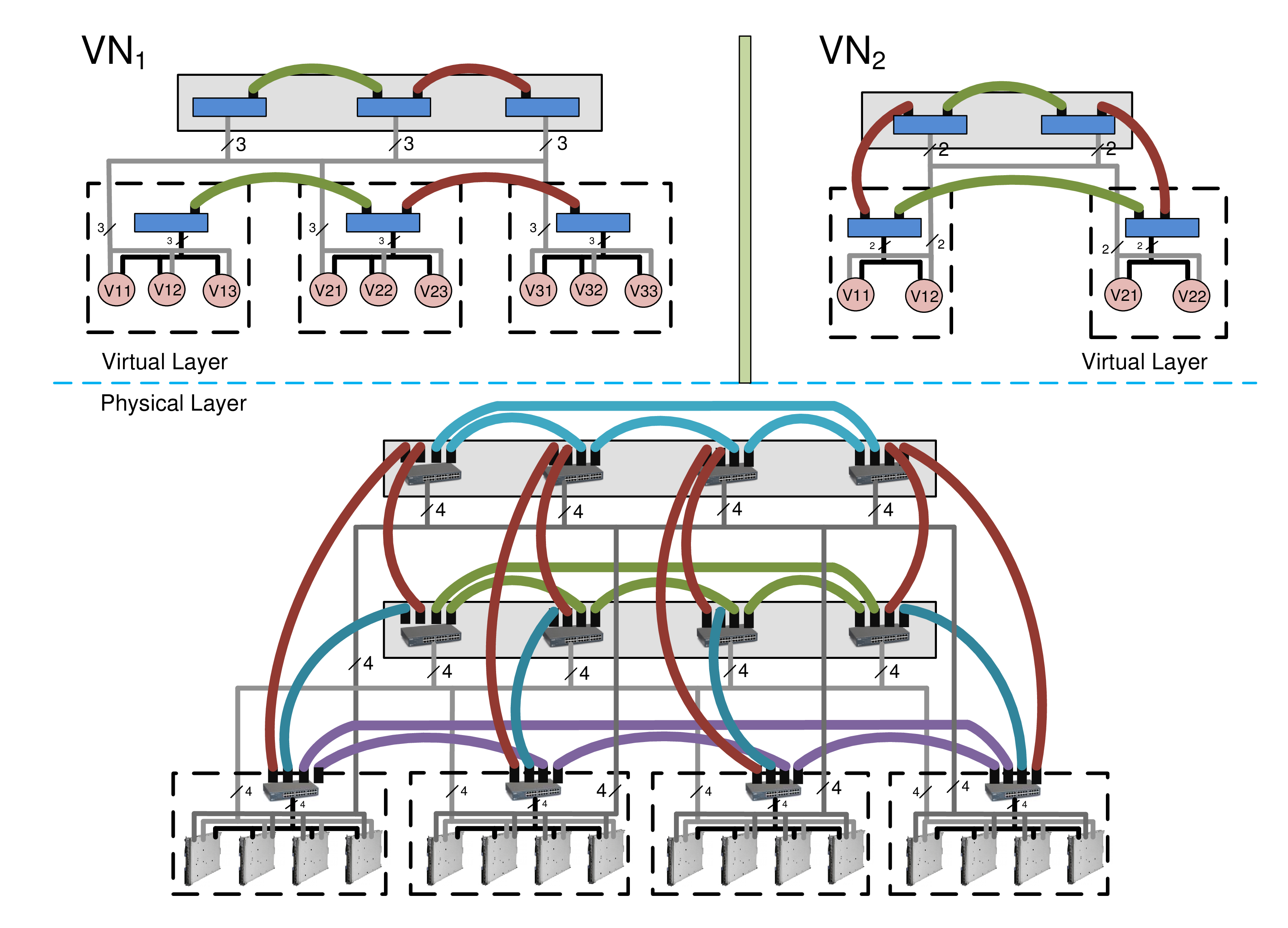} \\
(a) & (b)
\end{tabular}
\caption{A schematic illustration of the ToD approach.}
\label{Fig-Topoloy_on-demand-schematic}
\end{figure}

In contrast, we propose to generalize the interactive topology-matching concept in order to handle temporal changes in the required bandwidth. In particular, in this approach that we call Topology on-Demand (ToD), the VNet communicates its specific new requirements in terms of location and bandwidth to the SeP/InP, and the InP proposes a new topology that could handle the new traffic requirements. In particular, the specific topology with certain features such as IBW is provided when and where the VNet users needs it. In other words, the topology can be dynamically changed when the VNet needs it, and it is maintained in an on-demand style.
To be more specific, in the ToD approach, the VNet does not need to be aware of the structure of the physical network. Instead, at the moment of a high traffic spark, a {\em ToD middleware} would initiate a dynamic change request without requiring any explicit action from the client of the application except determining the bandwidth requirements.
The ToD requests could also be initiated by the applications themselves prior to their high-BW actions.
A schematic illustration of the ToD approach is presented in Fig. \ref{Fig-Topoloy_on-demand-schematic}, where two VNets are served using a Hybrid-BCube physical topology. Fig. \ref{Fig-Topoloy_on-demand-schematic}-(a) shows the {\em status quo} before the ToD requests, and an anticipated response of the ToD middleware to changes in traffic requirements of the VNets is provided in Fig. \ref{Fig-Topoloy_on-demand-schematic}-(b).  These concepts will be explored in details in the future.

\section{Conclusion and future prospects}
\label{sec:conclusion}
This paper investigated a group of DC network topologies with dynamic bandwidth-efficiency and failure-resiliency deliverable on demand based on the active traffic in a virtualized DC network. 
A classic BCube topology is considered as a topology benchmark throughout the paper. 
In particular, three main approaches were proposed to modify the structure of a classic BCube topology toward improving their associated structural features and maximum achievable interconnected bandwidth for both cases of single- and multi-path routing scenarios.
Finally, we run an extensive simulation program to compare the performance of the proposed topologies when failure of components and also traffic congestion are presented. 
Our simulation experiments showed the efficiency of the proposed modified topologies compared to the benchmark in terms of the bandwidth availability and failure resiliency.
We finally introduced some benefits of the proposed topologies in the design of the smart DC networks.
In particular, two new concepts of topology matching and topology on-demand in a virtual environment were introduced. These concepts could considerably reduce the overhead of virtual networks, and could also bring dynamic adaptability to variations in network requirements in real-time.
\bibliographystyle{IEEEtran}
\bibliography{imagepw}
\end{document}